\documentclass[11pt]{article}

\usepackage{epsfig}

\def\e{{\epsilon}}
\def\phi{{\varphi}}
\def\tPhi{{\tilde\Phi}}
\def\tphi{{\tilde\varphi}}
\def\mem{{\rm m}}
\def\ben{{\rm b}}

\def\eg{{\it e.g.}}
\def\ie{{\it i.e.}}

\title{Folding Energetics in Thin-Film Diaphragms}
\author{Gustavo Gioia$^{\rm a}$, Antonio DeSimone$^{\rm b}$,
  Michael Ortiz$^{\rm c}$, and  Alberto M.\ Cuiti\~no$^{\rm d}$ }  
\begin{document}
\date{Urbana, April 30, 2000}
\maketitle
\vskip -1cm
{\noindent \scriptsize ${}^{\rm a}$ 
Department of Theoretical \& Applied Mechanics,
University of Illinois,
Urbana, IL 61801, USA.}\\ 
 {\noindent \scriptsize ${}^{\rm b}$ 
 Max Planck Institute for Mathematics in the Sciences, 
 04103 Leipzig, Germany.}\\
 {\noindent \scriptsize ${}^{\rm c}$ 
 Graduate Aeronautical Laboratories, 
California Institute of Technology, Pasadena, CA 91125, USA.} \\ 
{\noindent \scriptsize ${}^{\rm d}$ 
Department of Mechanical \& Aerospace Engineering,           
Rutgers University,
Piscataway, NJ 08854, USA.}

{\bf \noindent 
   We perform experiments on thin-film diaphragms 
  to show that the 
 folding patterns of anisotropically compressed diaphragms  
  are strikingly different from those of isotropically 
  compressed ones. We then use a simple  von K\'arm\'an 
 model to relate the overall features of these  
   folding patterns to the underlying energetics. We  
 show that the differences between the isotropic and 
 anisotropic cases can be traced back to 
   fundamental changes in the energy structure of the 
    diaphragms.    
  Finally, we point out that the energy structure of
 thin-film diaphragms  is similar to that of many other systems 
 in physics and engineering, into which our study may provide 
 interesting insights. 
 }

\begin{figure}
\epsfxsize=4.96in
\centerline{\epsfbox{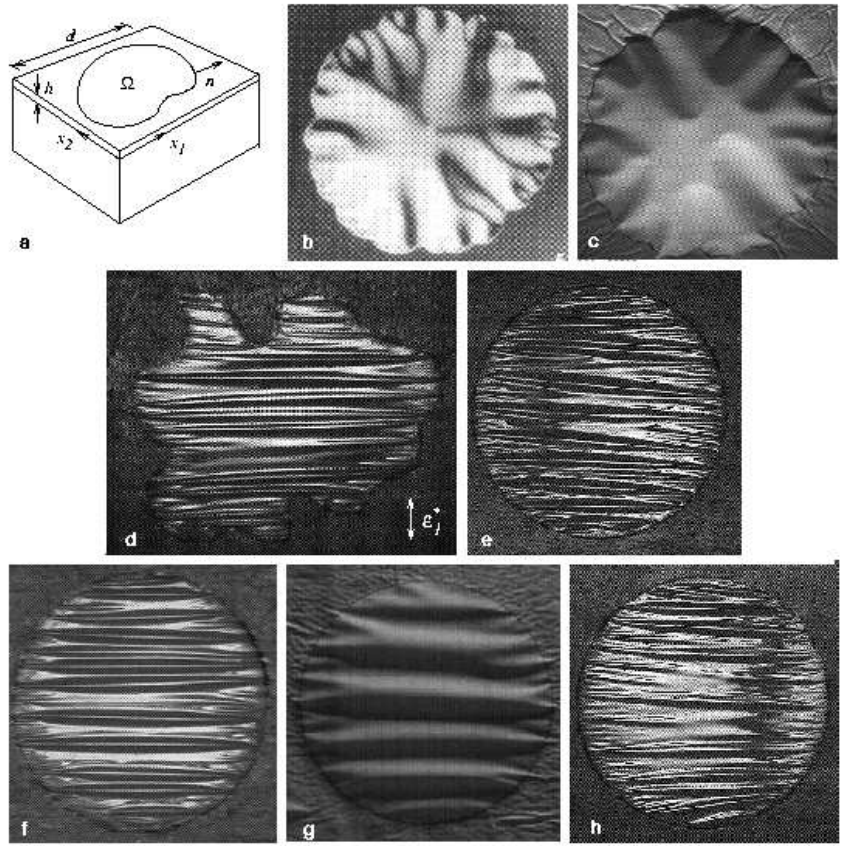}}
\caption{ {\bf a}, Thin film/substrate system, diaphragm, and notation.
 Top view of  folded diaphragms: {\bf b}, SiC film, 
 $\e^*_1=\e^*_2=0.011$,  $d/h=140$ [22]. 
 {\bf c}, Paper film, $\e^*_1=\e^*_2=0.040$,  $d/h=950$. 
 The following are polymeric films (except when noted)
  with $\e^*_2=0$: 
  {\bf d}, $\e^*_1=0.038$, $d/h=5600$. {\bf e}, $\e^*_1=0.035$,  $d/h=5600$.  
 {\bf f}, $\e^*_1=0.035$, $d/h=2840$. {\bf g}, Paper film, 
  $\e^*_1=0.040$  $d/h=1800$.  
 {\bf h}, $\e^*_1=0.019$, $d/h=5600$.
  }
\label{dia}
\vskip 2in
\end{figure}

\noindent Consider a film of thickness $h$ bonded to the flat surface of a  
 substrate except over a domain $\Omega$ of characteristic size $d$,
 Fig.~\ref{dia}(a). The portion $\Omega$ of the film is 
 a {\it diaphragm\/} [1].   If we apply a strain
 $\e^*$ to the substrate, in the plane of the film,  the
 diaphragm may   deflect out of the substrate and fold [2].
 We study the conditions which $\e^*$ must fulfill for this to occur, 
 and the nature of the resulting folding. 

 We prepared  diaphragms by gluing  paper sheets and 
 polymeric films ($h\sim 0.01$ to $0.1\,$mm) onto  substrates 
 (thickness $\sim 30\,$mm) of the shape  shown in Fig.~\ref{dia}(a),
 made of a high-density styrofoam [3]. Then we compressed  
 the substrates  in two perpendicular directions, $x_1$ and $x_2$, 
using screw-driven steel plates. When the applied strain
 components are equal to each other,  $\e^*_1=\e^*_2$, the strain 
is  {\it isotropic\/}; otherwise it is  
 {\it anisotropic\/}. Figs.~\ref{dia}(b-c)  
 show the folding of two equally shaped diaphragms  subject to 
isotropic strains. The folding pattern is the same in both diaphragms; 
 the number and spatial arrangement of the folds  depend
 exclusively on the shape of $\Omega$, being independent of 
 $d/h$ and  of the strain.  A completely different situation 
obtains when the  strains are anisotropic.
 The folds are then perpendicular to the direction 
 of $\e^*_1$ (where $\e^*_1 >\e^*_2$ by convention),
 regardless of the shape of $\Omega$, Figs.~\ref{dia}(d-e).
 The number of folds depends on $d/h$, however, see 
 Figs.~\ref{dia}(e-g); a comparison of Figs.~\ref{dia}(e) and 
 (h) reveals that it depends also on the strain. 

 For the analysis of these results we follow [2] in 
 {\it i\/}) modeling the  film as a von K\'arm\'an plate, 
 and {\it ii\/}) constraining the in-plane displacements of 
 the diaphragm to remain null. As we shall show, this simple model 
 allows for a straighforward energetic interpretation of
   the overall characteristics of folding patterns  in thin-film 
 diaphragms; the study of specific features of the folding, 
 as well as of folding in structures other than diaphragms,  
  may require the consideration of the in-plane displacements [4]. 
  For the proposed model the bending 
   energy density at a point $(x,_1,x,_2{})$  of the diaphragm is 
\begin{equation}
\label{wben}
\phi^\ben = {{C h^2}\over{12}}[(1-\nu)(w,_{11}^2+w,_{22}^2+2\, w,_{12}^2)+\nu (w,_{11}
 + w,_{22})^2],
\end{equation}
and the membrane energy density is 
\begin{equation}
\label{wmem}
\phi^\mem={{C}\over{2}}\left[{{1}\over{4}}
 \left( |\nabla w|^2- 2(\e^*_1+\nu\e^*_2{}) \right)^2+ 
      (1-\nu) w,_2^2 (\e^*_1-\e^*_2{}) \right]+ \phi^\mem_{\rm min},
\end{equation}
where $C = E h/(1-\nu^2)$ is the membrane stiffness of the film;
  $E$ is the Young  modulus  and $\nu$ the Poisson ratio 
 of the film; $w$ is the out-of-plane deflection of the diaphragm; 
  $\nabla w=(w,_1,w,_2{})$ is the gradient of $w$; 
   $|\nabla w|=(w,_1^2+w,_2^2)^{1/2}$ is the largest slope
  of the diaphragm at the given point; and we have defined 
 $\phi^\mem_{\rm min}= C (1-\nu^2) (\e^*_2)^2/2$. The total 
energy of the diaphragm is the sum of the 
 bending and membrane energies, $\Phi=\Phi^\mem+\Phi^\ben$, 
 where $\Phi^\mem$ and $\Phi^\ben$ are the integrals over $\Omega$ of 
 $\phi^\mem$ and $\phi^\ben$, respectively.  We search a 
 folding, described by $w(x_1,x_2{})$, 
 which  minimizes $\Phi[w]$ subject to the 
 conditions $w = 0$ and  $w,_n= 0$ on the boundary, 
 Fig.~\ref{dia}(a).

We start by noting that $\Phi^\ben$ and 
  $\Phi^\mem$ are of order $h^3$ and $h$, respectively. 
 Since we are interested in the  thin-film limit,
 $h\rightarrow 0$, we expect $\Phi^\mem$ to admit one or
 more minimizers with the  same overall structure as the 
 minimizer of   $\Phi$. We know that in the minimizers of
$\Phi^\mem$ the  folds will not be rounded, but 
  take the form of lines of slope discontinuity or {\it sharp folds\/}. 
 (The reason is that $\Phi^\mem$ contains first derivatives 
 of $w$ only.)  The bending energy will be confined to 
 these sharp folds,  in the form of an energy per 
 unit length of sharp fold [5,6].  We plan to {\it i\/})
 find the foldings which minimize $\Phi^\mem$, and then {\it ii\/})
 select among them the one which contains the least
 sharp-fold energy; this one we shall call the {\it preferred  
 folding\/} [2,7,8].

 We try to minimize $\Phi^\mem$ by minimizing its integrand,  
 $\phi^\mem$, Eq.~\ref{wmem}.
   For convenience we define a  {\it  compressive regime\/} 
   $\e^*_1+ \nu \e^*_2 >0$, and a {\it characteristic slope\/}
 $k=\sqrt{\e^*_1+ \nu \e^*_2}$. We look for gradients 
 $\nabla w=(w,_1, w,_2{})$ for which $\phi^\mem$
  attains a minimum.  When  $\e^*$ falls  
 outside  the compressive regime, a single minimum 
 exists for $\nabla w=(0,0)$, and the diaphragm remains flat. 
  In the  isotropic case,  $\e^*_1=\e^*_2$, 
 infinitely many minima of value $\phi^\mem_{\rm min}$ 
 exist in the compressive regime; they occur 
  for $|\nabla w|=k$, \ie\ whenever the largest slope equals
 the characteristic slope. In the  anisotropic case, two minima 
 of value $\phi^\mem_{\rm min}$ 
 exist in the compressive regime; 
 they occur when $\nabla w=(\pm k,0)$.
 Thus for a diaphragm in the compressive regime  the 
 minimum possible of the membrane energy is 
 $\Phi^\mem_{\rm inf}= \phi^\mem_{\rm min} A_\Omega$, 
  where $A_\Omega$ is the area of the diaphragm. 
  For convenience we shall work 
 with $\tphi^\mem= \phi^\mem-\phi^\mem_{\rm min}$,
 which implies  $\tPhi^\mem=\Phi^\mem-\Phi^\mem_{\rm inf}$,  
 and  $\tPhi^\mem_{\rm inf}=0$. 

\begin{figure}
\epsfxsize=2.8in
\centerline{\epsfbox{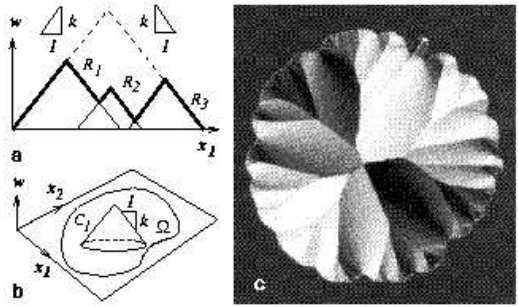}}
\caption{ Analysis of the isotropic case.  {\bf a}, Side view of a 1D, 
 constrained  diaphragm covered with a set of  three 
 simple roofs of slopes $\pm k$, the associated  minimizer (bold lines),  
 and  the preferred folding (dashed lines). {\bf b}, A 2D diaphragm 
 and an example of conical roof of slope $k$, denoted 
 $C_1$. {\bf c} Preferred folding for the diaphragm shape of 
 Figs.~\ref{dia}(b) and (c). }
\label{iso}
\end{figure}

 We study the isotropic case first.  Consider a 1D 
 example in which we  impose the constraint  $w,_2=0$. Then,
 the minima of  $\tphi^\mem$ occur for $w,_1=\pm k$, 
 and we can construct a minimizer of $\tPhi^\mem$ by covering  
 $\Omega$ with any set of {\it simple roofs\/} of slopes $\pm k$ 
 (such as $R_1$, $R_2$ and $R_3$ in Fig.~\ref{iso}(a)),  and then choosing 
 their upper envelope. Infinitely many  such minimizers exist, all of 
 which contain sharp folds, as expected. To minimize the energy 
 associated with the sharp folds, we must minimize their number; this we 
 effect by selecting  the upper envelope of all the minimizers, 
  Fig.~\ref{iso}(a).  In  2D a preferred folding can be found analogously, 
 as the upper envelope of all the possible cobertures of $\Omega$ with 
 conical roofs of slope $k$, Fig.~\ref{iso}(b) [2,9].
 The preferred folding 
 so selected does {\it i\/}) minimize $\Phi^\mem$;  {\it ii\/}) depend 
 exclusively on the shape of $\Omega$; and {\it iii\/})  match the 
 experimental observations, as  shown by a comparison of
 Fig.~\ref{iso}(c) with  Figs.~\ref{dia}(b) and (c).

\begin{figure}
\epsfxsize=4.8in
\centerline{\epsfbox{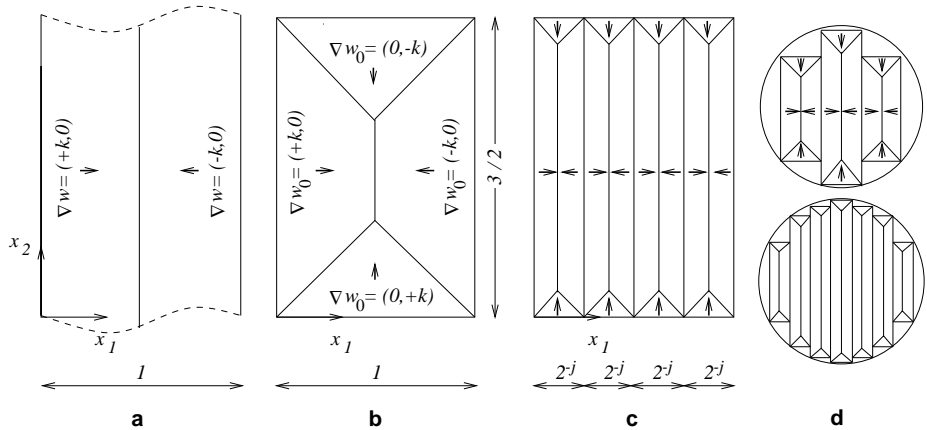}}
\caption{ Analysis of the anisotropic case.  {\bf a}, Top view of 
   a semi-infinite diaphragm with preferred folding.
 {\bf b}, Rectangular diaphragm with folding $w_0$.  
  {\bf c}, Idem with $w_2$ (\ie\ the microstructure $j=2$ of the minimizing sequence
  $w_j$). {\bf d}, Two terms of a minimizing sequence for a circular diaphragm. }
\label{aniso}
\end{figure}

 Now we turn to the anisotropic case. Consider first the infinite
 diaphragm of Fig.~\ref{aniso}(a). For this diaphragm the preferred 
 folding can be found as in the 1D isotropic case, 
 in the form of a single simple roof  of slopes $w,_1=\pm k$, 
 Fig.~\ref{aniso}(a).   
 In the finite diaphragm of Fig.~\ref{aniso}(b), on the other hand,
 a single simple roof violates the  boundary  conditions 
 $w(x_2=0)=w(x_2=3/2)=0$. To circumvent this problem we 
 try the folding $w_0$ of Fig.~\ref{aniso}(b). On the triangular
  regions of  Fig.~\ref{aniso}(b), which we call {\it closure domains\/},
 $w,_1\neq\pm k$, and therefore   
 $\tphi^\mem_\triangle > \tphi^\mem_{\rm min}$.
  It follows that $w_0$ is not a minimizer. In fact,   
 calling  $A_\triangle[w_0]$ the area of the 
 closure domains, $\tPhi^\mem[w_0]= \tphi^\mem_\triangle A_\triangle{}[w_0] >
  \tPhi^\mem_{\rm inf}$. We can approach the minimum value   
 $\tPhi^\mem_{\rm inf}=0$, however, by using foldings $w_j$ with  
 $j=1,2,\ldots\,$, \eg\ Fig.~\ref{aniso}C, 
  for which $A_\triangle[w_j]\rightarrow 0$ and therefore 
  $\tPhi^\mem[w_j]=\tphi^\mem_\triangle A_\triangle[w_j]\rightarrow 0$ 
  as $j\rightarrow \infty$.  Thus {\it $\tPhi^\mem$ can be made 
  arbitrarily close to its minimum  by  allowing the diaphragm to become 
 highly folded.\/} The sequence $w_j$ is said to be a 
 {\it minimizing sequence}, and the associated foldings are 
 called {\it microstructures\/} [10,11].

 For arbitrarily shaped diaphragms,  we can construct 
 minimizing sequences in an analogous way, \eg\   Fig.~\ref{aniso}(d).
  All the microstructures in a minimizing sequence contain sharp folds 
 and closure domains. As the microstructures become more folded,
  the sharp-fold energy increases, and the closure-domain 
 energy decreases. In principle, we can identify a 
  preferred microstructure for which 
   the trade off between closure-domain energy and sharp-fold energy  
  is resolved in the least total energy. Thus bending checks the infinite
  folding implied by a minimizing sequence [10]. A straightforward 
 dimensional analysis reveals that the number of folds $N$ in the 
 preferred microstructure scales with $\tphi^\mem_\triangle d/T$,
 where $T$ is the energy per unit length of sharp fold. 
  Setting $\e^*_2=0$ for the sake of simplicity, it is easy to conclude
 that $\tphi^\mem_\triangle\propto E (\e^*_1)^2 h$; a more involved analysis 
 leads to $T\propto E (\e^*_1)^{3/2} h^2$ [12]. Therefore, 
\begin{equation}
\label{ene}
 N\propto {{d}\over{h}} (\e^*_1)^{1/2}.
\end{equation}
 According to this expression, the number of folds
 for the diaphragms of  Figs.~\ref{dia}(d-h) should be proportional
 to 32, 30, 16, 10 and 22, respectively. The observed numbers 
 are 32, 28, 24, 12, and 22 (folds spanning the whole diaphragm).
   We conclude that the preferred microstructure of the minimizing sequence 
 does {\it i\/}) fail to minimize $\Phi^\mem$; and {\it ii\/})  
  contain, in accord with the experimental evidence, 
  folds which are perpendicular to the direction of $\e^*_1$,
  regardless  of the shape of $\Omega$, in a number 
  which scales with $d/h$ and $\e^*$ in the form of Eq.~(\ref{ene}).

 To sum up, the folding of  compressed diaphragms can be 
 interpreted in terms of two operations: 
 fold to release membrane energy, and then allow bending to 
 select one among many possible foldings.  
  In the isotropic case, infinitely many foldings exist
 which can  accomodate the boundary  conditions and simultaneously minimize
  $\Phi^\mem$. Out of these foldings bending selects a preferred
  one. In the anisotropic case, no folding 
  exists which can accomodate the boundary conditions and 
 simultaneously minimize $\Phi^\mem$.  It is possible, however, to 
 construct sequences of increasingly  fine foldings, or microstructures, 
  whose associated membrane energies converge 
  to the minimum value of $\Phi^\mem$.  Out of these foldings 
 bending selects a preferred one. 

The occurence of microstructures in
  compressed diaphragms discloses a mathematical similarity 
   with many problems 
 including  solidification [13], 
 solid-state phase transformation [14], 
 epitaxial thin-film growth [15], crystal plasticity [16], 
 ferromagnetism [17],  {\it et cetera\/},
  where microstructures have been documented in the  form    
  of eutectic structures, twinning, film roughening, 
  dislocation cells, and magnetic domain structures, respectively. 
 Other such problems are coagulation [18], stretching  
  of solid foams [19],  
  and self-assembly of polymer layers on patterned 
  substrates [20].   Because experiments can be easily performed on 
 compressed diaphragms, and a simple  model
 appears to explain their behavior well, the study of 
  diaphragms may prove useful to gain insights into 
  many  other systems.
 For example, a 90 degree rotation of the gradients in 
 Figs.~\ref{aniso}(b-d) 
 leads to closed-flux vector fields such as they obtain
 in ferromagnets, where magnetic poles are energetically
  penalized [17].
  In this analogy, sharp folds model
  ferromagnetic domain walls, and strain anisotropy models
  crystalline anisotropy.  A comparison of the fold
  branching observable close to  
 diaphragm boundaries, \eg\ Fig.~\ref{dia}(g),
 with the analogous phenomenon of domain branching 
  in the vicinity of free surfaces in ferromagnets [21]
 would enhance our understanding of the underlying energetics.
  Supported by a grant from the 
 Mechanics and Structures of 
  Materials Program, NSF, Dr.\ K.\ P.\ Chong, Program Director.
 
\vskip .6cm 

\subsection*{References and notes}
\setlength{\itemsep}{0.ex}
\frenchspacing        
\begin{enumerate}

\item Thin-film diaphragms have elicited much interest in 
  recent years because of their applications in micro 
  electro-mechanical systems; see almost any chapter in 
  M.\ Madou,  {\it Fundamentals of   Microfabrication} 
 (CRC Press, Boca Raton, Florida, 1997). 

\item G.\ Gioia, G., and  M.\ Ortiz, 
        {\it Adv.\ Appl.\ Mech.\/} {\bf 33}, 119 (1997).

\item A relatively high density ($\sim 40\,$kg/m$^3$)
 is required for the foam 
 to deform homogeneously. In fact, low-density solid foams
  display  deformation patterns which are analogous to 
  the folding patterns of anisotropically compressed diaphragms; 
 see  also [19].

\item A.\ Lobkovsky, S.\ Gentges, H.\ Li, D.\ Morse, and T.\ Witten,
     {\it Science\/} {\bf 270}, 1482  (1995); E.\ Cerda, 
 S.\ Chaieb, F.\ Melo, and L.\ Mahadevan, 
      {\it Nature\/} {\bf 401}, 46  (1999). 

\item  L.\ Modica,
 {\it Arch.\ Rat.\ Mech.\ Anal.\/}  {\bf 98}, 123 (1987). 

\item  R.\ V.\ Kohn, and S.\ M\"uller,  
   {\it Comm.\ Pure Appl.\ Math.\/}, {\bf 47}, 405 (1994).

\item E.\ de Giorgi, 
  {\it Rendiconti di Matematica\/} {\bf 8}, 277 (1975).

\item  P.\ Sternberg,  
 {\it Arch.\ Rat.\ Mech.\ Anal.\/} {\bf 101}, 
   209 (1988). 

\item W.\ Jin, {\it Singular Perturbation and the Energy of Folds\/}, 
   PhD Thesis, Courant Institute of Mathematical 
   Sciences, New York University (1997).

\item  J.\ M.\ Ball,   and R.\ D.\  James, 
 {\it Arch.\ Ration.\ Mech.\ Anal.\/} {\bf 100}, 13 (1987).

\item R.\ V.\ Kohn, 
  {\it Continuum Mech.\ Thermodyn.\/} {\bf 3}, 193 (1991).

\item G.\ Gioia, A.\ DeSimone,  and A.\ M.\ Cuiti\~no 
   (to be published).

\item  J.\ W.\  Cahn, 
  {\it Acta Metall.\/} {\bf 9}, 795 (1961). 

\item  A.\ G.\ Kachaturyan,  
 {\it  Theory of 
  Structural Transformations 
 in Solids.\/} (J.\ Willey \& Sons, New York, 1983).

\item  C.\ Orme,  and B.\ G.\ Orr, 
{\it Surf.\ Rev.\ Lett.\/} 
        {\bf 4}, 71 (1997).

\item  M.\ Ortiz, and  E.\ A.\ Repetto,  
   {\it J.\ Mech.\ Phys.\ Solids\/} {\bf 47}, 397 (1999).

\item  A.\ DeSimone,
  {\it Arch.\ Ration.\ Mech.\ Anal.\/} {\bf 125}, 99 (1993).

\item  J.\ Carr and R.\ Pego, 
 {\it Proc.\ Roy.\ Soc.\ London\/}, 
   A{\bf 436},  569 (1992). 


\item  Y.\ Wang, G.\ Gioia, A.\ M.\ and Cuiti\~no (to be published).

\item   M.\ B\"oltau, S.\ Walheim,  J.\ Mlynek, G.\ Krausch, 
   and U.\ Steiner,
   {\it Nature\/} {\bf 391}, 877 (1998).

\item  R.\ V.\ Kohn and S.\ M\"uller, 
  {\it Phil.\ Mag.\/} A{\bf 66}, 697 (1992).

\item  A.\ S.\ Argon,  V.\ Gupta, H.\ S.\ Landis,  and J.\ A.\ Cornie, 
 {\it J.\ Mater.\ Sci.\/}  {\bf 24}, 1207 (1989).

\end{enumerate}

\noindent  \vskip .4cm 

 
\newpage

\end{document}